# Fermi pockets and quantum oscillations in specific heat of YBCO in the presence of disorder

Partha Goswami[1], Manju Rani[1], and Avinashi Kapoor[2]

[1] D.B.College, University of Delhi, Kalkaji, New Delhi-110019, India

[2] Department of Electronic Science, University of Delhi south campus, New Delhi-110021, India

**Abstract** We investigate a chiral d-density wave (CDDW) mean field model Hamiltonian in the momentum space suitable for the hole-doped cuprates, such as YBCO, in the pseudo-gap phase to obtain the Fermi surface(FS)topologies, including the anisotropy parameter($\acute{\varepsilon}$) and the elastic scattering by disorder potential ($|v_0|$). For $\acute{\varepsilon} = 0$, the chemical potential $\mu = -0.27$ eV for 10% doping level, and $|v_0| \geq |t|$ (where $|t| = 0.25$ eV is the first neighbor hopping), at zero/non-zero magnetic field (B), the FS on the first Brillouin zone are found to correspond to Fermi pockets around anti-nodal regions and barely visible patches around nodal regions. For $\acute{\varepsilon} \neq 0$, we find Pomeranchuk distortion of FS. We next relate our findings regarding FS to the magneto-quantum oscillations in the electronic specific heat. Since the nodal quasi-particle energy values for B = 0 are found to be greater than μ for $|v_0| \geq |t|$, the origin of the oscillations for non-zero B corresponds to the Fermi pockets around anti-nodal regions. The oscillations are shown to take place for 17 T ≤ B ≤ 53 T in the weak disorder regime ($|v_0|$=0.25eV) only.

**1 Introduction** In the past ten years, extensive theoretical [1,2,3,4,5,6,7,8.9] and experimental efforts [10, 11,13,14, 15,16,17,18,19] have been devoted to understand the pseudogap (PG) phenomenon in the normal state of the underdoped cuprates. The intriguing aspect of the PG phase of the hole doped cuprates, such as YBCO, is that it could be characterized with a variety of coexisting/competing orders as have been reported by previous workers[1,2,3,4,5,6,7,8,9,10,11,12]. Their enterprises have led to understanding of the Fermi surface (FS) topologies of the cuprates without and with magnetic field (B) background. The theorists [1,2,3,4,5,6,7,8.9] among them have put forward explanations of the (experimentally) observed facts/anomalies [11,13,14,15, 16,17,18,19,20] in various physical properties, which carry the signature of the complexities in PG state, with high degree of success. In this communication, starting with a chiral d-density wave (CDDW)[1,2] mean field Hamiltonian $H_{d+id}$ in momentum space involving suitable dispersion with hopping integrals ($t_{ij}$), and incorporating the effect of elastic scattering by disorder potential ($|v_0|$) in the Fourier coefficient of single-particle Green's function ( defined by $H_{d+id}$ ), we show that the Fermi pockets in the anti-nodal regions and patches in the nodal regions of FS, at zero/non-zero magnetic field, can be obtained from the energy eigen-values of the matrix in $H_{d+id}$ for $|v_0| \geq |t|$ where t is the first-neighbor hopping integral. It may be mentioned that the "algebraic charge liquid" picture of Senthil et al. [9] predicts two kinds of hole-like pockets, viz. the elliptic and the banana-shaped. Our finding in section 3 is inconsistent with the latter. The angle resolved photo-emission spectroscpic (ARPES) studies[21,22,23, 24,25,26] (including the vacuum ultra-violet (VUV) laser-based ARPES [27]) of cuprates, where the experimental observations roughly correspond to the so-called "maximal intensity surface" explained in ref.[23], have not shown the evidence of the existence of the Fermi pockets so far. Taillefer et al.[14,15,16], however, have detected quantum oscillations in the electrical resistance of under-doped YBCO establishing the existence of a well-defined FS with Fermi pockets in the anti-nodal region when the superconductivity is suppressed by a magnetic field. Furthermore, Boebinger et al.[13] have observed the oscillations in the specific heat of YBCO-Ortho II samples(in the presence of a magnetic field B = 45 T), the same type YBCO samples investigated by Taillefer and co-workers [14,15,16], confirming their findings. The motivations behind our theoretical work are (i) to show the existence of Fermi pockets (FP) in the anti-nodal regions, and (ii) to explore the possibility of approximate (1/B)-oscillations in the specific heat in an effort to verify the finding of Boebinger et al.[13]. While we succeed in showing FPs (see Figs.2 and 3), the quantum oscillations are shown to be possible for 17 T ≤ B ≤ 53 T in the weak disorder regime ($|v_0|$ = 0.25 eV) only.

The reason for the identification of the PG state with the CDDW state rather than the well-known [3,4] d-density wave (DDW) state is that the CDDW ordering offers a theoretical explanation [5] of the non-zero polar Kerr effect observed recently in YBCO by Kapitulnik et al.[11]. Also, since the experimental signature of nematic order has been observed recently in cuprates in neutron scattering experiments [10], it is felt that the hopping anisotropy [28], which captures the electronic nematicity at the mean-field level, ought to be a part of the ongoing investigation to be dealt with at a deeper level in future.

The paper is organized as follows: In section 2 we present the mean field Hamiltonian $H_{d+id}$ in momentum space involving in-plane hopping anisotropy. In section 3 we discuss the elastic scattering by impurities and relate it to the issue of the Fermi pockets as this occupies the centre-stage [29] in the magneto-quantum oscillation context. In section 4 we discuss the issue of the magneto-quantum oscillations in electronic specific heat in detail. The paper ends in section 5 with the concluding remarks.

**2 Model Hamiltonian** For a magnetic field applied in z-direction, we consider a normal state tight-binding energy dispersion involving t, t´, t´´ which are, respectively, the hopping elements between nearest, next-nearest (NN) and NNN neighbours:

$$\varepsilon_{k,N}(B) = -2(t_x \cos(k_x a) + t_y \cos(k_y a + \varphi))$$
$$+ 4t´ \cos(k_x a) \cos(k_y a + \varphi/2) - 2(t_x´´\cos 2k_x a$$
$$+ t_y´´\cos 2(k_y a + \varphi)) + \hbar(N + (1/2))\omega_c. \quad (1)$$

Here the Landau level(LL) index N = 0, 1... and 'a' is the lattice constant (of YBCO). The vector potential **A** is assumed to be in the Landau gauge: **A**= (0 −Bx 0). The quantity $\omega_c = eB/m^*$ is the cyclotron frequency where $m^*$ is the effective mass of the electrons. The Zeeman splitting has not been taken into account. Following Hackl and Vojta [28], we have introduced a hopping anisotropy parameter έ, such that the hopping elements obey $t_{x,y}= (1 \pm έ/2)t$ and $t´´_{x,y} = (1 \pm έ/2)t´´$. For έ ≠ 0, the lattice rotation symmetry is spontaneously broken. In the numerical calculations below we take t as an energy unit where t = 0.25 eV, t´ = 0.4t, t´´= 0.0444 t. These values are the same as in ref.[3]. Furthermore, we shall consider the value of the chemical potential( μ) of the fermion number to be (– 0.27 eV) for 10 % hole doping (ref.[3]). In the presence of a vector potential **A**, the hopping amplitudes $t_{ij}$, corresponding to the sites **i** and **j**, assume the form $[t_{ij}\exp(a_{ij})]$ where $a_{ij} = (\pi/\Phi_0) \int_j^i \mathbf{A} \cdot \mathbf{dl}$ and $\Phi_0 = (h/2e)$. For the first neighbor hopping, say, corresponding to the sites **i** = (a,0) and **j** = (a,a), the quantity $a_{ij} = -(\pi/\Phi_0) \int_{(a,a)}^{(a,0)} Bxdy = \varphi$ where $\varphi = (2\pi eBa^2/h)$ is the Peierls phase factor. Similarly, for the second neighbor hopping, say, corresponding to the sites **i** = (a,0) and **j** = (2a,a), the quantity $a_{ij} = -(\pi/\Phi_0) \int_{(2a,a)}^{(a,0)} Bydy = \varphi/2$. These explain the reason behind the appearance of $\varphi$ and $\varphi/2$, respectively, in the first and the second terms of Eq.(1). Liewise, the reason for the appearance of $\varphi$ in the third term of Eq.(1) could be explained.

In the second-quantized notation, the Hamiltonian (with index j = (1,2) below corresponding to two layers of YBCO) for the Chiral d+id density-wave state, together with the anisotropy in the hopping parameters, in the presence of magnetic field (B) can be expressed as

$$H_{d+id}(B) = \sum_{k,N,\sigma,i=1,2} \Phi^{(i)\dagger}_{k,\sigma} E_N(k,B) \Phi^{(i)}_{k,\sigma} \quad (2)$$

where $\Phi^{(i)\dagger}_{k,\sigma} = (d^{\dagger(1)}_{k,\sigma} \ d^{\dagger(1)}_{k+Q,\sigma} \ d^{(2)\dagger}_{k,\sigma} \ d^{\dagger(2)}_{k+Q,\sigma})$ and $E_N(k,B) = [\varepsilon_{k,N}^U(B) I_{4\times 4} + \zeta_k(B) \cdot \boldsymbol{\alpha}]$. Here $\boldsymbol{\alpha} = (\alpha_1 \ \alpha_2 \ \alpha_3 \ \alpha_4)$ with

$$\alpha_i = \begin{pmatrix} \sigma_i & 0 \\ 0 & \sigma_i \end{pmatrix} (i=1,2,3), \ \alpha_4 = \begin{pmatrix} 0 & I_{2\times 2} \\ I_{2\times 2} & 0 \end{pmatrix}. \quad (3)$$

$I_{4\times 4}$ and $I_{2\times 2}$, respectively, are the 4×4 and 2×2 unit matrices; $\sigma_i$ are the Pauli matrices and $\zeta_k(B) = (-\chi_k \ -\Delta_k \ \varepsilon_k^L(B) \ t_k)$ – a four-component vector. Here the chiral order parameter [1], $D_k \exp(i\theta_k)$, is given by $D_k = (\chi_k^2 + \Delta_k^2)^{1/2}$ and $\cot\theta_k = (-\chi_k/\Delta_k)$ with

$$\chi_k = -(\chi_0/2)\sin(k_x a)\sin(k_y a), \quad (4)$$

and
$$\Delta_k = (\Delta_0(T)/2)(\cos k_x a - \cos k_y a). \quad (5)$$

We consider the simplest form of the modulation vector, **Q** = (±π, ±π). The quantity $t_k$ is momentum conserving tunneling matrix element which for the tetragonal structure of cuprates is given by $t_k = (t_0/4)(\cos k_x a - \cos k_y a)^2$. As in ref.[3], we take $t_0 = 0.032 t$. The reader may note that here we have disregarded the LL mixing completely. The energy eigenvalues of $E_N(k,B)$ are

$$E_N^{(j,\nu)}(k,B) = [\varepsilon_{k,N}^U(B) + j w_{k,N}(B) + \nu t_k] \quad (6)$$

where $\varepsilon_{k,N}^U(B) = (\varepsilon_{k,N}(B) + \varepsilon_{k+Q,N}(B))/2$, $\varepsilon_{k,N}^L(B) = (\varepsilon_{k,N}(B) - \varepsilon_{k+Q,N}(B))/2$, and $w_{k,N}(B) = [(\varepsilon_{k,N}^L(B))^2 + D_k^2]^{1/2}$. Here j is equal to (± 1) with j = +1 corresponding to the upper branch (U) and j = −1 to the lower branch(L); for a given j, ν = ± 1. For LL index N = 0,1, we shall have two-fold splitting for a given (j,ν). With these eigenvalues, we obtain the non-interacting Matsubara propagator

$$G_N(k,\omega_n) = \sum_{\nu = \pm 1} \{ V_{k,N}^{(U,\nu)2}(i\omega_n - E_N^{(U,\nu)}(k,B))^{-1}$$
$$+ V_{k,N}^{(L,\nu)2}(i\omega_n - E_N^{(L,\nu)}(k,B))^{-1} \}. \quad (7)$$

The quasi-particle coherence factors $(V_{k,N}^{(U,\nu)2}, V_{k,N}^{(L,\nu)2})$ are given by the expressions

$$V_{k,N}^{(U,\nu)2} = (1/4)[1 + (\varepsilon_{k,N}^L/w_{k,N})], \quad (8)$$

$$V_{k,N}^{(L,\nu)2} = (1/4)[1 - (\varepsilon_{k,N}^L/w_{k,N})]. \quad (9)$$

The magnetic field dependence of these factors arise through $\varepsilon_{k,N}^L(B)$. At hole doping level ~ 10%, we find that the pseudo-gap(PG) temperature T* ~ 155 K. We have assumed, the value $\Delta_0(T < T^*) = 0.0825$ eV $= 0.3300$ t in the vicinity of T*, and $(\chi_0/\Delta_0(T<T^*))^2 = 0.0025$. We shall now consider the effect of the elastic scattering by impurities on the Fermi surface topology and search for the evidence of the existence of Fermi pockets. This is an important issue as without these pockets the Onsager relation [29] does not allow one to investigate magneto-quantum oscillations.

The impurity potential/disorder with finite range not only has drastic effects on the Fermi surface (FS) topology, but will be seen to affect the density of states at Fermi energy relevant for thermodynamic properties in the following section.

**3. Elastic scattering by impurities and Fermi energy density of states** The effect of elastic scattering by impurities involves the calculation of self-energy $\Sigma(\mathbf{k},\omega_n)$, in terms of the momentum and the Matsubara frequencies $\omega_n$, which alters the single-particle excitation spectrum in a fundamental way. A few diagrams contributing to the self-energy are shown in Fig.1. The wiggly lines carry momentum but no energy as the scattering is assumed to be elastic. The total momentum entering each impurity vertex, depicted by a slim ellipse, is zero. We assume that impurities are alike, distributed randomly, and contribute a potential term $V(\mathbf{r} = \mathbf{R} + z\,\mathbf{k}) = \sum_i u(\mathbf{r} - \mathbf{r}_i)$ where $u(\mathbf{r} - \mathbf{r}_i)$ is the potential due to a single impurity at $\mathbf{r}_i = \mathbf{R}_i + z\,\mathbf{k}$ for a given $z$ and $\mathbf{R} = x\,\mathbf{i} + y\,\mathbf{j}$. The potential term $u(\mathbf{r} - \mathbf{r}_i)$ is expanded in a Fourier series $V(\mathbf{r}) = \sum_{\mathbf{q},i} V(|\mathbf{q}|)\exp[i\mathbf{q}\cdot(\mathbf{r} - \mathbf{r}_i)]$. We first consider only the contribution of the Fig.1(a). Assuming the scattering by impurities weak, as in ref.[30], we may write it as $\Sigma^{(1)}(\mathbf{k},\omega_n) = N_j \sum_{\mathbf{k}'} |V(\mathbf{k}-\mathbf{k}')|^2 G_N(\mathbf{k}',\omega_n) = \Sigma_0^{(1)}(\mathbf{k},\omega_n) + \Sigma_e$ where

$$\Sigma_0^{(1)}(\mathbf{k},\omega_n) = -N_j \sum_{\mathbf{k}', \nu = \pm 1} |V(\mathbf{k}-\mathbf{k}')|^2 (i\omega_n) \int_{-\infty}^{+\infty} d\varepsilon\, \rho(\varepsilon)$$
$$\times [\, V_{k,N}^{(U,\nu)2}\,(\omega_n^2 + E_N^{(U,\nu)}(k,B)^2)^{-1}$$
$$+ V_{k,N}^{(L,\nu)2}\,(\omega_n^2 + E_N^{(L,\nu)}(k,B)^2)^{-1}], \quad (10)$$

$N_j$ is the impurity concentration, $V(\mathbf{k}-\mathbf{k}')$ characterizes the momentum dependent impurity potential, and $\Sigma_e$ is the part of the first order contribution which can be shown to be independent of $\mathbf{k}$ and $\omega_n$ for $\mathbf{k}$ close to Fermi momentum. We take $\Sigma_e = 0.05$ eV below. To evaluate the integrals in (10), such as $\int_{-\infty}^{+\infty} d\varepsilon\, \rho(\varepsilon)\,(\omega_n^2 + E_N^{(j,\nu)}(k,B)^2)^{-1}$, we assume $\rho(\varepsilon) = \rho_0 \delta(\varepsilon - E_N^{(j,\nu)}(k,B))$ where we take a broad band-width, say, $(10t) \sim 2.5$ eV which gives $\rho_0 \sim 0.4$ (eV)$^{-1}$. We thus obtain

$$G_N(\mathbf{k},\omega_n) \approx -\rho_0 (i\omega_n)(\pi/|\omega_n|), \quad (11a)$$

$$\Sigma_0^{(1)}(\mathbf{k},\omega_n) = -N_j\,\rho_0\,(i\omega_n)\sum_{\mathbf{k}'}|V(\mathbf{k}-\mathbf{k}')|^2 (\pi/|\omega_n|). \quad (11b)$$

We may write the right-hand side of Eq.(11) as $[-i\omega_n/(2|\omega_n|\tau_k)]$, where $(1/\tau_k) = 2\pi N_j\,\rho_0 \sum_{\mathbf{k}'}|V(\mathbf{k}-\mathbf{k}')|^2$. Note that $\tau_k$, which corresponds to quasi-particle lifetime (QPLT), is expressed in reciprocal energy units. Upon using the Dyson's equation, the full propagator may be written as $G^{(Full)}_N(\mathbf{k},\omega_n) \approx \sum_{\nu = \pm 1}\{D/E\}$, where

$$D = [V_{k,N}^{(U,\nu)2}(i\omega_n - E_N^{(L,\nu)}) + V_{k,N}^{(L,\nu)2}(i\omega_n - E_N^{(U,\nu)})],$$

$$E = (i\omega_n)^2 - (i\omega_n)(E_N^{(U,\nu)} + E_N^{(L,\nu)} - (i/2\tau_k) + \Sigma_e)$$
$$+ \{E_N^{(U,\nu)} E_N^{(L,\nu)} + (-(i/2\tau_k) + \Sigma_e)$$
$$\times (V_{k,N}^{(U,\nu)2} E_N^{(L,\nu)} + V_{k,N}^{(L,\nu)2} E_N^{(U,\nu)})\}. \quad (12)$$

We have dropped the argument part from the single-particle excitation spectra above for convenience. The roots of the equation $E = 0$ are

$$i\omega_n = \{(\,E_N^{(U,\nu)} + E_N^{(L,\nu)} - (i/2\tau_k) + \Sigma_e)/2\}$$
$$\pm (R^{(\nu)}_{k,N}{}^{1/2}/2)(\cos(\theta^{(\nu)}_{k,N}/2) - i\sin(\theta^{(\nu)}_{k,N}/2)),$$

$R^{(\nu)}_{k,N} = [E_1^{(\nu)2} + E_2^{(\nu)2}]^{1/2},\; \tan(\theta^{(\nu)}_{k,N}) = E_2^{(\nu)}/E_1^{(\nu)},$

$E_1^{(\nu)} = (E_N^{(U,\nu)} - E_N^{(L,\nu)})^2 + \Sigma_e^2 - (1/4\tau_k^2) + 2(E_N^{(U,\nu)} + E_N^{(L,\nu)})\Sigma_e$
$$- 4\Sigma_e(V_{k,N}^{(U,\nu)2} E_N^{(L,\nu)} + V_{k,N}^{(L,\nu)2} E_N^{(U,\nu)}),$$

$E_2^{(\nu)} = \{(E_N^{(U,\nu)} + E_N^{(L,\nu)} + \Sigma_e)/\tau_k\}$
$$- (2/\tau_k)(V_{k,N}^{(U,\nu)2} E_N^{(L,\nu)} + V_{k,N}^{(L,\nu)2} E_N^{(U,\nu)}). \quad (13)$$

It follows that the denominator $E$ of $G^{(Full)}_N(\mathbf{k},\omega_n)$ may be written as the product of two factors $(i\omega_n - (\alpha_{\nu,N}^{(+)} + i\beta_{\nu,N}^{(+)}))(i\omega_n - (\alpha_{\nu,N}^{(-)} + i\beta_{\nu,N}^{(-)}))$ where

$$\alpha_{\nu,N}^{(j=\pm)} = \{(E_N^{(U,\nu)} + E_N^{(L,\nu)} + \Sigma_e)/2\}$$
$$\pm (R^{(\nu)}_{k,N}{}^{1/2}/2)\cos(\theta^{(\nu)}_{k,N}/2),$$

$$\beta_{\nu,N}^{(j=\pm)} = -\{(1/4\tau_k) \pm (R^{(\nu)}_{k,N}{}^{1/2}/2)\sin(\theta^{(\nu)}_{k,N}/2)\}. \quad (14)$$

In view of (14), $G^{(Full)}_N(\mathbf{k},\omega_n) \approx \sum_{\nu = \pm 1}\{D/E\}$ may be written as

$G^{(Full)}_N(\mathbf{k},\omega_n) = \sum_{\nu=\pm 1} V_{ren,k,N}^{(+,\nu)2}[i\omega_n - \acute{\varepsilon}_r^{(+,\nu)}{}_N + i(1/4\tau_{k,N}^{(+,\nu)})]^{-1}$
$$+ V_{ren,k,N}^{(-,\nu)2}[i\omega_n - \acute{\varepsilon}_r^{(-,\nu)}{}_N + i(1/4\tau_{k,\alpha,N}^{(-,\nu)})]^{-1} \quad (15)$$

where the superscript $j = (\pm)$,

$$\acute{\varepsilon}_r^{(j,\nu)}{}_N = \alpha_{\nu,N}^{(j)},$$

$$(1/\tau_{k,N}^{(j,\nu)}) = \{(1/\tau_k) \pm (2R^{(\nu)}_{k,N}{}^{1/2})\sin(\theta^{(\nu)}_{k,N}/2)\},$$

$$V_{ren,k,N}^{(j,\nu)2} = (1/4)(1\pm\delta_{k,N}),\; \delta_{k,N} = (\delta^{(1)}_{k,N}/\delta^{(2)}_{k,N}), \quad (16)$$

and

$\delta^{(1)}_{k,\alpha,N} = [(\alpha_{\nu,N}^{(+)} + i\beta_{\nu,N}^{(+)}) + (\alpha_{\nu,N}^{(-)} + i\beta_{\nu,N}^{(-)})$
$$-2(V_{k,N}^{(+,\nu)2} E_N^{(-,\nu)} + V_{k,N}^{(-,\nu)2} E_N^{(+,\nu)})], \quad (17)$$

$$\delta^{(2)}_{k,N} = [(\alpha_{\nu,N}^{(+)} + i\beta_{\nu,N}^{(+)}) - (\alpha_{\nu,N}^{(-)} + i\beta_{\nu,N}^{(-)})] \quad (18)$$

The renormalized Bogoluibov coherence factors $V_{ren,k,N}^{(j,\nu)2}$ turn out to be complex quantities.

We have calculated explicitly the propagators $G^{(Full)}_N(\mathbf{k},\omega_n)$ above with the inclusion of impurity scattering. The corres-ponding retarded Green's function $G^{(R)}_N(\mathbf{k},t)$, in units such that $\hbar=1$, is given by $G^{(R)}_N(\mathbf{k},t) = {}_{-\infty}\!\int^{+\infty}(d\omega/2\pi)\exp(-i\omega t)\, G^{(Full)}_N(\mathbf{k},\omega)$ where in the upper and lower half-plane, respectively,

$$G^{(Full)}_N(\mathbf{k},\omega)=\sum_{j=(\pm),v=\pm 1} V_{ren,k,N}^{(j,v)2}[i\omega_n-\acute{\varepsilon}_r^{(j,v)}{}_N+(i/4\tau_{k,N}^{(j,v)})]^{-1}$$
(19)

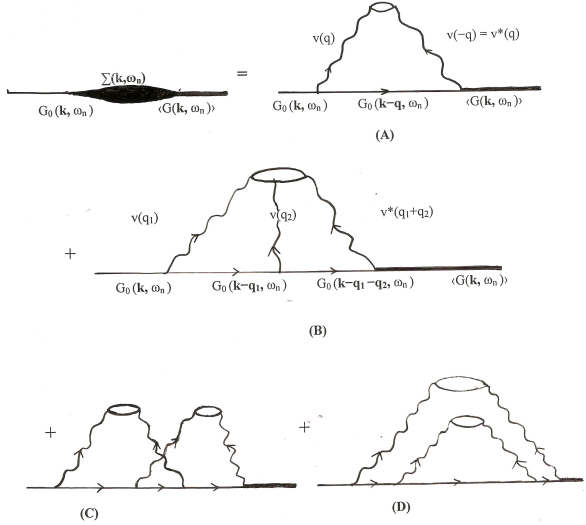

**Figure 1** A few diagrams contributing to the self-energy. The wiggly lines carry momentum but no energy. The total momentum entering each impurity vertex, depicted by a slim ellipse, is zero. We have assumed that impurities are alike, and distributed randomly. Whereas Figs.(A) and (B) correspond to one impurity vertex, the Figs.(C) and (D) correspond to a product of four impurity potentials with non-zero averages. These are the cases where two impurities each give rise to two potentials. Thus the figures involve the interference of the scattering by more than one impurity. We have assumed low concentration of impurities and therefore these figures yield smaller contributions compared to those corresponding to 1(A), 1(B) and the other diagrams of the same class involving only one impurity vertex.

and

$$G^{(Full)}_N(\mathbf{k},\omega)=\sum_{j=(\pm),v=\pm 1} V_{ren,k,N}^{(j,v)2}[i\omega_n-\acute{\varepsilon}_r^{(j,v)}{}_N-(i/4\tau_{k,N}^{(j,v)})]^{-1}.$$
(20)

Thus $G^{(R)}_N(\mathbf{k},\omega') = {}_{-\infty}\!\int^{+\infty}dt \exp(i\omega' t)\, G^{(R)}_N(\mathbf{k},t)$ is given by (19) with $\omega$ real. We obtain

$$G^{(R)}_N(\mathbf{k},t) = \sum_{j=(\pm),v=\pm 1} V_{ren,k,N}^{(j,v)2}\, i\exp(-i\,\acute{\varepsilon}_r^{(j,v)}{}_N(\mathbf{k})t$$
$$-(t/4\tau_{k,N}^{(j,v)}))\,\theta(t) \quad (21)$$

where the unit step function $\theta(t)={}_{-\infty}\!\int^{+\infty}(id\omega/2\pi)\{\exp(-i\omega t)/(\omega+i\,0^+)\}$. The electronic excitations in cuprates are thus demonstrably Bogoliubov quasi-particles in the pseudo-gap phase with finite lifetime for the states of definite momentum due to the impurity scattering. Using the integral representation of $\theta(t)$ above it is not difficult to show that

$$G^{(R)}_N(\mathbf{k},\omega') = {}_{-\infty}\!\int^{+\infty}dt\exp(i\omega' t)\, G^{(R)}_N(\mathbf{k},t)$$
$$=\sum_{j=(\pm),v=\pm 1} V_{ren,k,N}^{(j,v)2}\,[\omega-\acute{\varepsilon}_r^{(j,v)}{}_N+(i/4\tau_{k,N}^{(j,v)})]$$
$$\times[(\omega-\acute{\varepsilon}_r^{(j,v)}{}_N)^2+(1/4\tau_{k,N}^{(j,v)})^2]^{-1}. \quad (22)$$

As the renormalized Bogoluibov coherence factors $V_{ren,k,N}^{(j,v)2}$ are found to be complex, the dimensionless density of states $\rho_N(\mathbf{k},\omega) \equiv (-1/2\pi^2\rho_0)\,\mathrm{Im}G^{(R)}_N(\mathbf{k},\omega)$ comprises of two parts: $\rho_N(\mathbf{k},\omega)= \rho'_N(\mathbf{k},\omega)+\rho''_N(\mathbf{k},\omega)$, where

$$\rho'_N(\mathbf{k},\omega) = (1/2\pi^2\rho_0)\sum_{j=(\pm),v=\pm 1}(\mathrm{Re}\,V_{ren,k,N}^{(j,v)2})\,\gamma^{(j,v)}_{k,N}$$
$$\times[(\omega-\acute{\varepsilon}_r^{(j,v)}{}_N(\mathbf{k}))^2+\gamma^{(j,v)2}_{k,N}]^{-1}, \quad (23)$$

$$\rho''_{\alpha,N}(\mathbf{k},\omega) = (-1/\pi\rho_0)\sum_{j=(\pm),v=\pm 1}(\mathrm{Im}\,V_{ren,k,N}^{(j,v)2})$$
$$\times(\omega-\acute{\varepsilon}_r^{(j,v)}{}_N(\mathbf{k}))\times[(\omega-\acute{\varepsilon}_r^{(j,v)}{}_N(\mathbf{k}))^2+\gamma^{(j,v)2}_{k,N}]^{-1}, \quad (24)$$

$$\mathrm{Re}\,V_{ren,k,N}^{(j,v)2} = (1/4)(1\pm\mathrm{Re}\,\delta_{k,\alpha,N}),$$
$$\mathrm{Im}\,V_{ren,k,N}^{(j,v)2} = (1/4)(1\pm\mathrm{Im}\,\delta_{k,\alpha,N}), \quad (25)$$

and $\gamma^{(j,v)}_{k,N}=\tau_{k,N}^{(j,v)\,-1}/4$ (the level-broadening factors). In order to determine the Fermi energy density of states (DOS) $\rho_{N,Fermi}(\mathbf{k})$, since we shall put $\omega=\mu$ in (23) and (24), it is clear that $\rho_{N,Fermi}(\mathbf{k}) = \sum_{j=(\pm),v=\pm 1}\rho^{(j,v)}_{N,Fermi}(\mathbf{k})$ where

$$\rho^{(j,v)}_{N,Fermi}(\mathbf{k}) = (1/2\pi^2\rho_0)(\mathrm{Re}\,V_{ren,k,N}^{(j,v)2})\,\gamma^{(j,v)}_{k,N}$$
$$\times[(\mu-\acute{\varepsilon}_r^{(j,v)}{}_N(\mathbf{k}))^2+\gamma^{(j,v)2}_{k,N}]^{-1}. \quad (26)$$

Equation (24) does not contribute here as the branches of the Fermi surface are given by $(\acute{\varepsilon}_r^{(j,v)}{}_N(\mathbf{k})-\mu)=0$. However, for $\omega\ne\mu$, definitely this equation will contribute towards the DOS. The chemical potential $\mu$, according to the Luttinger rule, is given by the equation

$$(1+p) = \int d(\mathbf{k}a)\sum_{j,v,N}\rho^{(j,v)}_{N,Fermi}(\mathbf{k})$$
$$\times(\exp(\beta(\acute{\varepsilon}_r^{(j,v)}{}_N(\mathbf{k},B)-\mu))+1)^{-1} \quad (27)$$

where p is the doping level, $\int d(\mathbf{k}a)\to {}_{-\pi}\!\int^{+\pi}(d(k_xa)/2\pi)\,{}_{-\pi}\!\int^{+\pi}(d(k_ya)/2\pi)$, and $\beta=(k_BT)^{-1}$. It may be mentioned in passing that $\rho_{N,Fermi}(\mathbf{k},B=0)$ roughly corresponds to the so-called "maximal intensity surface" [22,23] of the ARPES studies provided the momentum dependence of the level broadening factors are ignored.

We model V(|**k**−**k**′|) by a screened exponential falloff of the form V(|**k**−**k**′|) = [|v$_0$|$^2$ κ$^2$/{|**k**−**k**′|$^2$+ κ$^2$}]$^{1/2}$ to consider the effect of the in-plane impurities, where κ$^{-1}$ characterizes the range of the impurity potential. The limit κ >> |**k**−**k**′|, which corresponds to a point-like isotropic scattering potential characterizing the in-plane impurities, will only be considered here for simplicity. At this stage,

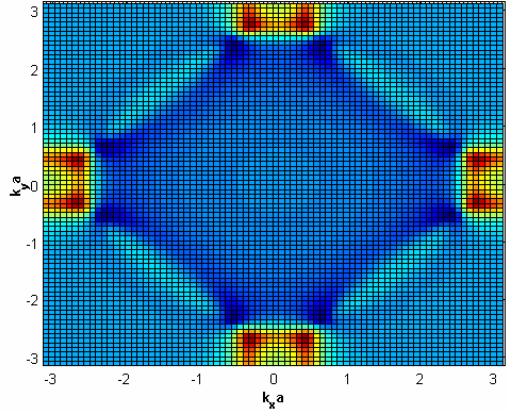

(a)

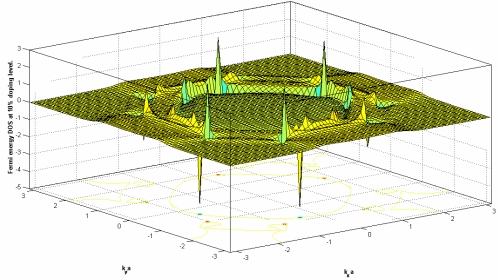

(b)

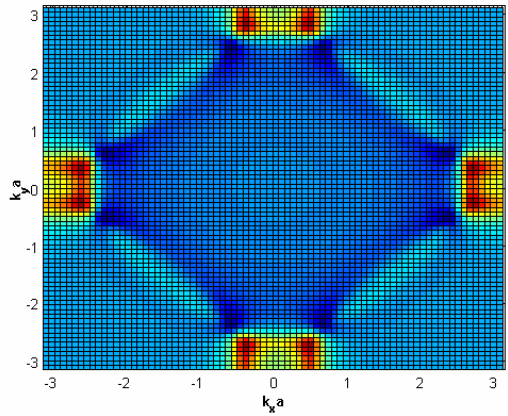

(c)

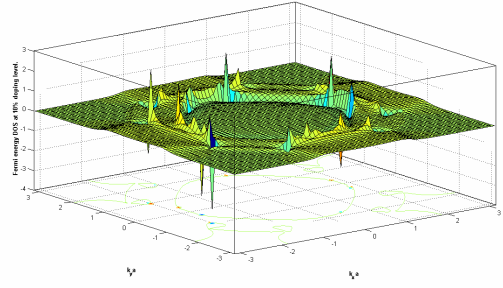

(d)

**Figure 2** The contour/3-D plots of the Fermi energy DOS on the Brillouin zone(BZ) at 10% hole doping, and B = 0. (a)The anisotropy parameter $\acute{\varepsilon}$ = 0 and the level broadening factor γ = 0.0357 eV.(b) $\acute{\varepsilon}$ =0 and γ = 0.2436 eV. (c) $\acute{\varepsilon}$ = 0.05 and γ = 0.0357eV. (d) $\acute{\varepsilon}$ = 0.05 and γ = 0.2436 eV. The scale of the plots in (a) and (c) is from −0.2 to 1. In the weak disorder regime we obtain prominent electron pockets and barely visible hole-like patches. In the strong disorder regime the scenario does not change radically. The comparison of Fig.(a) with (c)( and (b) with (d)) indicates the slight change (Pomeranchuk distortion) in the topology of the Fermi surface due to $\acute{\varepsilon}$ ≠ 0; electron pockets around (0,±π) widen sightly compared to those around (±π,0).

assuming low concentration of impurities, one may include the contributions of all such diagrams which involve only one impurity vertex. This gives the equation to determine the total self-energy $\Sigma_N(\mathbf{k},\omega_n)$:

$$\Sigma_N(\mathbf{k},\omega_n) = N_j\sum_q V(\mathbf{q}) G_N(\mathbf{k}-\mathbf{q},\omega_n) \Gamma_N(\mathbf{k},\mathbf{q},\omega_n) \quad (28)$$

where the Lippmann-Schwinger equation to determine $\Gamma_N(\mathbf{k},\mathbf{q},\omega_n)$ is

$$\Gamma_N(\mathbf{k},\mathbf{q},\omega_n) = V(-\mathbf{q})+\sum_{q'} V(\mathbf{q'}-\mathbf{q}) G_N(\mathbf{k}-\mathbf{q'},\omega_n)$$
$$\times \Gamma_N(\mathbf{k},\mathbf{q'},\omega_n). \quad (29)$$

This is t-martix approximation. Upon using the optical theorem for the t-matrix [30] one may write

$$\Sigma_N(\mathbf{k},\omega_n) = i \, \mathrm{Im} \, \Gamma_N(k,k,\omega_n) = -i\omega_n/(2|\omega_n|\acute{\Gamma}_{k,N}) \quad (30)$$

where $\acute{\Gamma}_{k,N}^{-1} = 2\pi N_j \rho_0 \sum_{k'}|\Gamma_N(\mathbf{k},\mathbf{k'})|^2$. Thus the effect of the inclusion of contibution of all the above mentioned diagrams is to replace the Born approximation for scattering by the exact scattering cross-section for a single impurity, i.e. $\tau_k^{-1} \rightarrow \acute{\Gamma}_k^{-1}$. Since $G_N(\mathbf{k},\omega_n)$ and $V(\mathbf{q})$ are known, using Eqs.(27), (28) and (29) one can determine $\acute{\Gamma}_{k,N}^{-1}$ in terms of $V(\mathbf{k})$. In the limit κ >> |**k**−**k**′|, the disorder potential $V(|\mathbf{q}|) \approx |v_0|$ and, therefore, from the latter we obtain

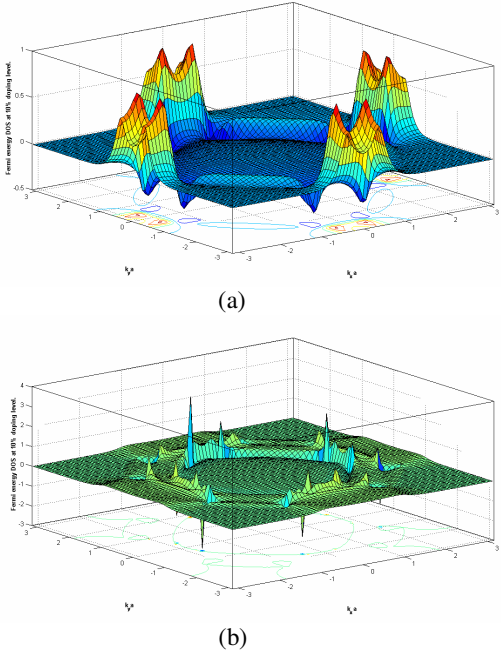

**Figure 3** The 3-D plots of the Fermi energy DOS on the Brillouin zone(BZ) at 10% hole doping and B = 45 Tesla. (a)The anisotropy parameter $\acute{\varepsilon} = 0$ and the level broadening factor $\gamma = 0.0357$ eV. (b) $\acute{\varepsilon} = 0$ and $\gamma = 0.2436$ eV. The features for B= 45 T are by and large the same as in Fig.2.

$$\Gamma_N(\mathbf{k},\omega_n) \approx |v_0|/(1 - |v_0| \ G_{\alpha,N}(\mathbf{k},\omega_n)) \quad (31)$$

In view of Eq.(11a), we find that

$$\text{Im } \Gamma_N(\mathbf{k},\omega_n) \approx - \rho_0 \pi |v_0|^2 /(1 + \rho_0^2 \pi^2 |v_0|^2). \quad (32)$$

From Eq.(30) we now find that $\acute{\Gamma}_{k,N}^{-1}$, in the first approximation, is given by $[2 \rho_0 \pi |v_0|^2 /(1 + \rho_0^2 \pi^2 |v_0|^2)]$. We take a moderate disorder potential $|v_0| = 0.25$ eV. This gives $\gamma_{\mathbf{k},N} = \acute{\Gamma}_{k,N}^{-1}/4 \approx 0.0357$ eV. We next take a stronger disorder potential $|v_0| = 1$ eV. This gives $\gamma_{\mathbf{k},N} = \acute{\Gamma}_{k,N}^{-1}/4 \approx 0.2436$ eV. We note that, even though $\acute{\Gamma}_{k,N}$ is found to be $\mathbf{k}$-independent in the first approximation, the term $\pm 4 R_{k,N}^{1/2} \sin(\theta_{k,N}/2)$ in (16) will ensure that $\tau_{k,N}^{(j,v)}$ are momentum dependent and different for the upper and lower branches. With these inputs we embark on a calculation of the specific heat in the next section.

**4 Magneto-quantum oscillations in specific heat** Following the Kadanoff-Baym approach[31], the thermodynamic potential may be given by the expression

$$\Omega(T,B,\mu) = \Omega_0(B) - 2(\beta N_s)^{-1}$$
$$\times \sum_{j,k,v,N} \ln\cosh(\beta(\acute{\varepsilon}_r^{(j,v)}{}_N(\mathbf{k},B) - \mu)/2) \quad (33)$$

where $\Omega_0(B) = N_s^{-1} \sum_{j,k,v,N}((\acute{\varepsilon}_r^{(j,v)}{}_N(\mathbf{k},B) - \mu)$ and $\beta = (k_B T)^{-1}$. The dimensionless entropy per unit cell is given by $S = \beta^2 (\partial\Omega/\partial\beta)$ while the electronic specific heat, for the pseudo-gapped (PG) phase (T < T*), is $C_{el} = -\beta (\partial S/\partial\beta)$. We obtain

$$C_{el} \approx 2 k_B N_s^{-1} \sum_{k,j,v,N} (\beta(\acute{\varepsilon}_r^{(j,v)}{}_N(\mathbf{k},B) - \mu))^2$$
$$\times \exp(\beta(\acute{\varepsilon}_r^{(j,v)}{}_N(\mathbf{k},B) - \mu)) \times (\exp(\beta(\acute{\varepsilon}_r^{(j,v)}{}_N(\mathbf{k},B) - \mu)) + 1)^{-2}. \quad (34)$$

We have ignored the temperature dependence of the chemical potential above. Now strictly speaking, for the magnetic field dependent phenomena, whenever we need to replace the sum over physical momenta $\mathbf{k}$ by an integral over a region of $\mathbf{k}$-space, the Berry-phase corrected result[35] is to be used. It is known[1,2,35] that for some isolated points in $\mathbf{k}$-space, the correction is much larger than those corresponding to the other points for a magnetic field B even of order 1 Tesla. In what follows, we, however, ignore this correction. Upon using (26) in (34) we find that the electronic specific heat at a given doping level is given by $C_{el} \approx \gamma(B) T$, where the specific heat coefficient for B $\neq 0$ may be expressed as

$$\gamma(B) \approx (k_B^2/\pi^2) \sum_{j,v,N} \int d(\mathbf{k}a) Q_1(B,\mathbf{k}) \times Q_2(B,\mathbf{k}), \quad (35)$$

$$Q_1(B,\mathbf{k}) = (\gamma^{(j,v)}{}_{\mathbf{k},N}/\rho_0) \times [(\mu - \acute{\varepsilon}_r^{(j,v)}{}_N(\mathbf{k}))^2 + \gamma^{(j,v)}{}_{\mathbf{k},N}^2]^{-1}, \quad (36)$$

$$Q_2(B,\mathbf{k}) = (\text{Re } V_{ren,k,N}^{(j,v)2}) \times \beta \times (\beta(\acute{\varepsilon}_r^{(j,v)}{}_N(\mathbf{k},B) - \mu))^2$$
$$\times \exp(\beta(\acute{\varepsilon}_r^{(j,v)}{}_N(\mathbf{k},B) - \mu)) \times (\exp(\beta(\acute{\varepsilon}_r^{(j,v)}{}_N(\mathbf{k},B) - \mu)) + 1)^{-2}. \quad (37)$$

The quantity $Q_2(B,\mathbf{k})$ is given by the expression

$$Q_2(B,\mathbf{k}) = \int_{-\infty}^{+\infty} dx \ I_{k,N}^{(j,v)}(x,B) \{x^2 e^x/(e^x + 1)^2\} \quad (38)$$

where
$$I_{k,N}^{(j,v)}(x,B) = (\hbar\omega_c)^{-1} (\text{Re } V_{ren,k,N}^{(j,v)2})$$
$$\times \delta((x/\beta\hbar\omega_c) - (\hbar\omega_c)^{-1} (\acute{\varepsilon}_r^{(j,v)}{}_N(\mathbf{k}, B) - \mu)). \quad (39)$$

The delta functions can be expanded in cosine Fourier series: $\delta(x-a) = (2\pi)^{-1} + \pi^{-1} \sum_{m=1}^{\infty} \cos[m(x - a)]$. Upon doing so, we find that the non-oscillatory part of the specific heat, with the linear dependence on T, is

$$C_{nonoscl} \sim (k_B^2 T/3\pi\hbar\omega_c) \sum_{j,v,N} \int d(\mathbf{k}a) Q_1(B,\mathbf{k}) \times (\text{Re } V_{ren,k,N}^{(j,v)2}).$$
$$(40)$$

The oscillatory part may be expressed as

$$C_{oscll} = (k_B^2 T/\pi^3 \hbar\omega_c) \sum_{j,v,N} \int d(\mathbf{k}a) Q_1(B,\mathbf{k}) \times (\text{Re } V_{ren,k,N}^{(j,v)2})$$
$$\times \int_{-\infty}^{+\infty} dx \sum_{m=1}^{\infty} \cos(mx/(\beta\hbar\omega_c)) \{x^2 e^x/(e^x + 1)^2\}$$

$$\times \cos\{m(\hbar\omega_c)^{-1}(\acute{\varepsilon}_r^{(j,v)}{}_N(\mathbf{k}, B) - \mu)\}. \quad (41)$$

Equation (41) prima facie indicates that the origin, of the approximate (1/B)-oscillations in the electronic specific heat of the under-doped YBCO, is the upper and the lower branches of the excitation spectrum. The appearance of the Dingle factors in (41), due to the scattering by impurities, is ensured by the Lorentzian $Q_1(B,\mathbf{k})$. In this Lorentzian, for the moderate disorder potential $|v_0| = 0.25$ eV, the level broadening factor(LBF) is 0.0357eV while, for the stronger disorder potential $|v_0| = 1$eV, the same is 0.2436 eV.

**Table 1** The values of the anti-nodal and the nodal excitations for B = 0 for the disorder potential $|v_0|$ = 0.25eV and 1 eV. All the values of $\tilde{E}_{n,N}(B=0)$ are above the value of $\mu = -0.27$ eV. Therefore, the origin of the specific heat oscillations shown in Eq.(41) would not correspond to the nodal patches around $\mathbf{k}=(\pm\pi/2,\pm\pi/2)$ shown in Fig.2.

| $|v_0|$ = 0.25eV | | $|v_0|$ = 1eV | |
|---|---|---|---|
| $\tilde{E}_{an,N}(B=0)$ in eV | $\tilde{E}_{n,N}(B=0)$ in eV | $\tilde{E}_{an,N}(B=0)$ In eV | $\tilde{E}_{n,N}(B=0)$ in eV |
| −0.2578 | 0.1639 | −0.0857 | 0.3389 |
| −0.5130* | 0.0249 | −0.6871* | −0.1501 |
| −0.2724** | 0.1639 | −0.0982 | 0.3389 |
| −0.5324* | 0.0249 | −0.7066* | −0.1501 |

* The energy values are less than $\mu = -0.27$ eV.
**The energy value is closest to $\mu = -0.27$ eV.

**Table 2** The values of the anti-nodal and the nodal excitations for B = 45 T for the disorder potential $|v_0|$ = 0.25eV and 1 eV.The first four rows correspond to LL index N = 0 while the next four to N = 1. All the values of $\tilde{E}_{n,N}(B= 45$ T) are above the value of $\mu = -0.27$ eV.

| $|v_0|$ = 0.25eV | | $|v_0|$ = 1eV | |
|---|---|---|---|
| $\tilde{E}_{an,N}(B=45$ T) in eV | $\tilde{E}_{n,N}(B=45$ T) in eV | $\tilde{E}_{an,N}(B=45$ T) in eV | $\tilde{E}_{n,N}(B=45$ T) in eV |
| −0.2578 | 0.1695 | −0.0837 | 0.3418 |
| −0.5098* | 0.0245 | −0.6839* | −0.1478 |
| −0.2703** | 0.1695 | −0.0962 | 0.3418 |
| −0.5293* | 0.0245 | −0.7034* | −0.1478 |
| −0.2538 | 0.1766 | −0.0797 | 0.3472 |
| −0.5035* | 0.0278 | −0.6775* | −0.1429 |
| −0.2662*** | 0.1766 | −0.0921 | 0.3472 |
| −0.5230* | 0.0278 | −0.6971* | −0.1429 |

* The energy values are less than $\mu = -0.27$ eV.
**This is the energy value (corresponding to LL index N = 0) which increases to −0.2700 eV at B = 53 T.
***This is the energy value (corresponding to N = 1) which attains the value of $\mu = -0.2700$ eV at B = 17 T.

In an effort to explain the possible quantum oscillations in the specific heat, for the anisotropy parameter $\acute{\varepsilon} = 0$ with LL index N=0 and 1, we introduce the two quantities $\tilde{E}_{an,N}$ and $\tilde{E}_{n,N}$, where

$$\acute{\varepsilon}_r^{(\pm,v)}{}_N(\mathbf{k}, B)\big|_{\mathbf{k}=[(\pm\pi,0),(0,\pm\pi)]} \equiv \tilde{E}_{an,N}(B) \quad (42)$$

and

$$\acute{\varepsilon}_r^{(\pm,v)}{}_N(\mathbf{k}, B)\big|_{\mathbf{k}=(\pm\pi/2,\pm\pi/2)} \equiv \tilde{E}_{n,N}(B), \quad (43)$$

corresponding to the anti-nodal and the nodal excitations respectively and calculate these quantities for the disorder potential $|v_0| = 0.25$eV and 1 eV. Each of the quantities $\tilde{E}_{an,N}(B)$ and $\tilde{E}_{n,N}(B)$ will have 8 values corresponding to the superscripts (±,v) and the subscript N. Since we take N = 0 and 1, the 4×4=16 possible values are displayed in Table 1 for B = 0; likewise, 8×4=32 possible values are in Table 2 for B = 45 T. From Tables 1 and 2(see also table legends below) we clearly notice that as the magnetic field(B) is increased, provided one is in the weak disorder regime ($|v_0| = 0.25$ eV), the Landau states move to a higher energy, ultimately rising above the Fermi level. They are thereby emptied, and the excess fermions find a place in the next lower Landau level (LL). During the crossing of a LL, its occupation by fermions is halted and then reduced. The specific heat consequently decreases slightly. As the excess fermions get accommodated in the next lower LL, the specific heat rises again. Thus the oscillations in the fermion density in the vicinity of Fermi energy manifest themselves as oscillations in the specific heat. These events take place for 17 T ≤ B ≤ 53 T (see the table legends). The similar oscillations in the electrical conductivity (Shubnikov-de Haas oscillations (SdHO)) is already reported [14,15,16]. As we notice above, in YBCO, these oscillations have their origin at the electron pockets in the Fermi surface located around the anti-nodal points. The key input in this analysis is the Landau level split dispersion in (1); the chirality aspect perhaps plays a minor role as these oscillations are also possible in the pure d-density wave state[4].

**5 Concluding remarks** The well-known theoretical developments, such as the dynamical mean field theory (DMFT)[7,8], the symmetry-constrained variational procedure of Wu et al. [31]etc., may require in future a revisit of the problem of the quantum oscillations with a new perspective. Particularly, the development of DMFT and its cluster extensions provide new path to investigate strongly correlated systems; the DMFT study of superconductivity near the Mott transition establishes the remarkable coexistence of a superconducting gap, stemming from the anomalous self-energy, with a pseudo-gap stemming from the normal self-energy. This theory also leads to the generation of the Fermi arc behavior of the spectral function [7,8].

In conclusion, we note that the computation of the correction to the quantum oscillations due to the Berry phase is an important future task. We note that the inclusion of the elastic scattering by impurities though has led to a clearer understanding, of the Fermi surface topology in the presence of a magnetic field at the semi-phenomenological level, the further examination of the single-particle excitation spectrum of the system in a fully self-consistent approximation framework is necessary to impart a comprehensive microscopic basis to the findings presented. Finally, we hope that our results, viz. the one relating to the reconstructed Fermi surface and the other to the electronic specific heat anomaly, will persuade researchers to look for them in the hole-doped cuprates. It must be added that the experimental observation of the latter is quite a difficult proposition, for the dominant phononic contribution is expected to overshadow the anomaly in the heat capacity measurements [32].